\documentclass[twocolumn,twoside,amsmath,amssymb,superscriptaddress,showpacs,nofootinbib,aps]{revtex4-1}

\usepackage{slashed}
\usepackage{amsmath}
\usepackage{amsthm}
\usepackage{subfigure}
\usepackage[utf8]{inputenc}
\usepackage{graphicx}
\usepackage{epsfig}
\usepackage{amssymb}
\usepackage{dcolumn}
\usepackage{bm}
\usepackage{color}
\usepackage[dvipsnames]{xcolor}

\newcommand{\ba}{\begin{array}}
\newcommand{\ea}{\end{array}}
\def\br{\begin{eqnarray}}
\def\er{\end{eqnarray}}
\def\be{\begin{equation}}
\def\ee{\end{equation}}
\usepackage{hyperref}

\def\({\left(}
\def\){\right)}

\def\<{\left\langle}
\def\>{\right\rangle}

\newcommand{\Q}{\textnormal{\tiny \textsc{Q}}}
\newcommand{\T}{\textnormal{\tiny \textsc{T}}}
\newcommand{\E}{\textnormal{\tiny \textsc{E}}}
\def\tt{\textnormal\tiny\textsc}

\begin{document}

\title{Technicolor models with coupled systems of Schwinger-Dyson equations}

\author{A. Doff}
\email{agomes@utfpr.edu.br}

\affiliation{Universidade Tecnol\'ogica Federal do Paran\'a - UTFPR - DAFIS
Av Monteiro Lobato Km 04, 84016-210, Ponta Grossa, PR, Brazil}

\author{A. A. Natale} 
\email{adriano.natale@unesp.br}

\affiliation{Instituto de F{\'i}sica Te\'orica - UNESP, Rua Dr. Bento T. Ferraz, 271,\\ Bloco II, 01140-070, S\~ao Paulo, SP, Brazil}

\begin{abstract}
When technicolor (TC), QCD, extended technicolor (ETC) and other interactions become coupled through their different Schwinger-Dyson equations, the solution of
these equations are modified compared to those of the isolated equations. The change in the self-energies is similar to that
obtained in the presence of four-fermion interactions, but without their \textit{ad hoc} inclusion in the theory. 
In this case TC and QCD self-energies decrease logarithmically with the momenta, which allows us to build models 
where ETC boson masses can be pushed to very high energies, and their effects will barely appear at present energies. 
Here we present a detailed discussion of this class of TC models. We first review the Schwinger-Dyson TC and QCD coupled
equations and explain the origin of the asymptotic self-energies. We develop the basic ideas of how viable TC models may be built
along this line, where ordinary lepton masses are naturally lighter than quark masses. One specific unified TC model 
associated with a necessary horizontal (or family) symmetry is described. The values of scalar and pseudo-Goldstone boson masses in this class
of models are also discussed, as well as the value of the trilinear scalar coupling, and the consistency of the models with the experimental constraints.
\end{abstract}



\maketitle

\section{introduction}

Over the years there have been many attempts to solve some of the drawbacks of the Standard Model (SM) related to the presence of
a fundamental scalar boson (like the hierarchy problem, triviality, etc...). Some of the proposals along these lines are interesting
due to the fact that fundamental scalar bosons fit naturally into these models, as in supersymmetric models~\cite{su1,su2,su3} 
and asymptotically safe SM extensions~\cite{sa00,sa11}. However, no signals of these theories have appeared so far. The Higgs particle found at the LHC~\cite{atlas,cms}
may be the first signal of a fundamental scalar boson, although the possibility that this boson is a composite one has not yet been discarded, and in this case some of the SM problems commented above may be alleviated.

Scalar bosons are essential to the mechanisms of chiral and gauge symmetry breaking in the SM, but it should be remembered that most
of what we have learned about the mechanisms of spontaneous symmetry breaking is based on the presence of composite or pair-correlated
scalar states, as happens in the Nambu-Jona-Lasinio model, QCD chiral symmetry breaking, and in the microscopic BCS theory of superconductivity.
For instance, chiral symmetry breaking is promoted in QCD by a nontrivial vacuum expectation value of a fermion
bilinear operator and the role of the Higgs boson is played by the composite $\sigma$ meson. These types of gauge theory models, dubbed technicolor (TC), were proposed 
40 years ago~\cite{wei,sus} and reviewed in Refs.~\cite{far,mira}. The many variations of these models continue to be studied~\cite{an,sa,sa1,sa2,sa3,sa4,be},
but no phenomenologically viable model has been found so far.

It is clear that building SM extensions in order to solve unknown questions (like the origin of the fermionic mass spectra), is easier when we
deal with fundamental scalar bosons, than when the spontaneous symmetry breaking
is promoted by composite scalars, even if we are far from solving the problems related to the existence of fundamental scalar bosons. The difficulty in models with
composite scalar boson resides in knowing the dynamics of the non-Abelian gauge theory responsible for their formation. 

We may say that the root of most TC problems lies in the way the ordinary fermions acquire their masses, which is
shown in Fig.1, where an ordinary fermion $f$ couples to a technifermion $F$ mediated by an extended technicolor (ETC) boson.
\begin{figure}[h]
\centering
\vspace*{-0.5cm}
\includegraphics[scale=0.45]{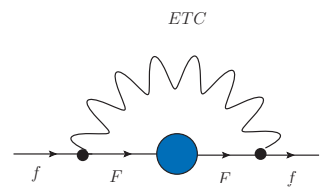} 
\vspace{-0.25cm} 
\caption[dummy0]{Ordinary fermion mass $f$ in ETC models}
\label{fig1}
\end{figure}
\noindent
\par  Assuming a standard non-Abelian TC self-energy ($\Sigma_{\T}$) given by~\cite{lane2}
\be
\Sigma_{\T} (p^2) \propto \frac{\mu_{\tt{TC}}^3}{p^2} \(\frac{p}{\mu_{\tt{TC}}}\)^{\gamma_m},
\label{eq1}
\ee
where $\mu_{\tt{TC}}$ is the characteristic TC dynamical mass at zero momentum (of order the Fermi mass) and $\gamma_m$ is the anomalous mass dimension (which depends on the TC coupling constant, and for an asymptotically free theory has a small value), the ordinary 
fermion mass turns out to be
\be
m_f \propto \frac{\mu_{\tt{TC}}^3}{M_{\E}^2},
\label{eq2}
\ee
where $M_{\E}$ is the ETC gauge boson mass. In order to explain the top-quark mass we need a small $M_{\E}$ value, and since ETC is one interaction that changes flavor, the simplest model that we can imagine will inevitably lead to flavor changing neutral currents incompatible with the experimental data (among other problems).

Solutions to the above dilemma seem to require a large  $\gamma_m$ value~\cite{holdom} leading to a TC self-energy with a harder momentum behavior,  and many models along these lines can be found in the literature~\cite{lane0,appel,yamawaki,aoki,appelquist,shro,kura,yama1,yama2,mira2,yama3,mira3,yama4}. In particular, we may quote the work of Takeuchi~\cite{takeuchi} where the TC Schwinger-Dyson equation (SDE) was solved with the introduction of an  four-fermion \textit{ad hoc} interaction, which can lead to the following expression for the TC self-energy:
\be
\Sigma_{\T}(p^2\rightarrow\infty)\propto \ln^{-\delta} (p^2/\mu_{\tt{TC}}^2),
\label{eq3}
\ee
where $\delta$ is a function of the many parameters of the model. The Takeuchi solution, when dominated by the four-fermion interaction,
is not different from the behavior of the self-energy when a bare mass is introduced into the theory, or from irregular SDE solution~\cite{lane2}.

\begin{figure}[h]
\centering
\includegraphics[scale=0.5]{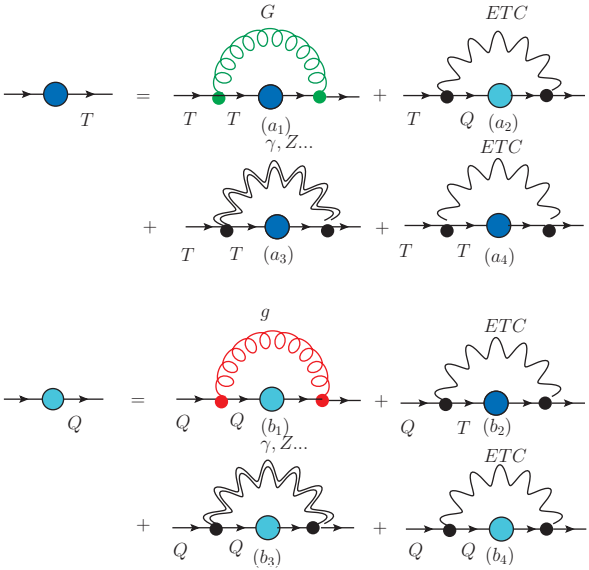} 
\vspace{-0.25cm} 
\caption[dummy0]{The coupled system of  SDEs for TC  ($T\equiv$technifermion) and QCD  ($Q\equiv$quark)  including  ETC and electroweak or other corrections. $G \,(g)$ indicates a technigluon (gluon).}
\label{fig2}
\end{figure}
Recently, we numerically solved the coupled TC [based on an $SU(2)$ group] and QCD gap equations~\cite{us1}, which are depicted in the Fig.2. It turned out that
both self-energies have the same asymptotic behavior as Eq.(\ref{eq3}). It is not difficult to understand the origin of such behavior. In Ref.~\cite{us2}
we analytically verified that the radiative corrections shown in Fig.2 act as an effective bare mass. In the case of ordinary quarks the second diagram ($b_2$)
on the right-hand of Fig.2  originates an effective mass due to TC condensation; on the other hand, the techniquarks obtain a tiny effective mass due to
QCD condensation [see diagram ($a_2$) in Fig.2], and an even larger mass due to the other diagrams [($a_3$) and ($a_4$)]. Therefore, the TC self-energy can be described by
\be
\Sigma_{\T}(p^2)\approx \mu_{\tt{TC}} \left[ 1+ \delta_1 \ln\left[(p^2+\mu^2_{\tt{TC}})/\mu^2_{\tt{TC}}\right] \right]^{-\delta_2} \,,
\label{eq4}
\ee
where $\delta_1$ and $\delta_2$ are parameters that will depend on the many possible SDE radiative corrections depicted in Fig.2; in particular, the dominant
correction to the technifermion masses will be generated by diagrams $(a_3)$ and $(a_4)$ of Fig.\ref{fig2}, and by diagram ($b_2$) in the case of ordinary fermion masses. We get a similar expression  for ordinary quarks, and it should be noticed that the \textit{isolated} infrared TC and QCD self-energy
behavior is the traditional one [the one associated to the regular solution or Eq.(\ref{eq1})] with dynamical masses of order $\mu_{\tt{TC}} \approx O(1)$TeV and 
$\mu_{\tt{QCD}}\approx 250$MeV, respectively, i.e., the coupled SDE system is a combination of the regular and irregular self-energy solutions~\cite{lane2}.
It is interesting to recall that such behavior is indeed that which minimizes the vacuum energy in gauge theories~\cite{us0}, and it is not different from Takeuchi's result
but rather originates from known interactions (QCD, for example).

The main consequence of the results of Refs. \cite{us1} and \cite{us2} [i.e., Eq.(\ref{eq4})] is that the dynamically generated masses will barely
depend on the ETC scale $M_\E$. In Ref.\cite{us1} we numerically verified that the ordinary quark masses behave as
\be
m_{\Q} \propto \lambda_E \mu_{\tt{TC}} [1+\kappa_1 \ln(M^2_{\E}/\mu_{\tt{TC}}^2)]^{-\kappa_2} \, ,
\label{eq5}
\ee
where $\lambda_E$ involves ETC couplings, a Casimir operator eigenvalue, and other constants, and $\kappa_i$ are related to the self-energies that enter in the calculation
of the generated masses, which is compatible with the quark mass computed with the help of Eq.(\ref{eq4}). Looking at Eq.(\ref{eq5}), it is clear that we can push the ETC scale up to the grand unification scale (or even the Planck scale) without large variations
of the $m_{\Q}$ values with $M_E$. It is also clear that the ordinary fermionic mass hierarchy will not arise from different $M_\E$ scales! The purpose of the present work it to
discuss how viable TC models can be built in this context, as well as to verify the phenomenological consequences of these models, and to show how that
they can be consistent with existing high-energy data.

It is important to note that the study of SDEs is very sophisticated, taking into account gluon-mass generation and possibly 
confinement~\cite{g1,g2,g3,g4,g5} as well as complex vertex structures~\cite{g6,g7}. However, the solutions discussed in Refs.~\cite{us1,us2} and in this work are related to the asymptotic behavior produced by the effective mass of the coupled SDE, and are not affected by the infrared intricacies of the strongly interacting theories.

The paper is organized as follows. In Sec. II we present one specific TC model, which is just an example of the many models that can be built
along the lines described in that section. We discuss the fact that a horizontal symmetry is necessary in this scheme. In Sec. III we discuss how a composite scalar boson can be lighter than the typical composition scale
of the theory responsible for this particular state. In Sec. IV we determine the order of magnitude of pseudo-Goldstone masses. In Sec. V
we compare the value of the TC condensate in our model with the one expected in walking TC theories. Section VI contains a
brief discussion of possible experimental consequences of the models discussed in Sec. II, and in Sec. VII we discuss what can be expected
regarding the trilinear scalar coupling. Section VIII contains our conclusions.

\section{Building TC models}

In Ref.~\cite{us1} we briefly proposed one specific TC model, which will be detailed here. As will be discussed at the end of this section, 
there is a large class of models that can be built along the 
same lines as the model described here. The model discussed in Ref.~\cite{us1} is based on the following group structure
$$
SU(9)_U \otimes SU(3)_H \,\, ,
$$
where the $SU(9)_U$ group is a non-Abelian grand unified theory (GUT) containing the SM and a $SU(4)_{\tt{TC}}$ group. The $SU(3)_H$ group is a horizontal or family
symmetry that is important for generating the hierarchy of fermion masses. 

There are several reasons for this particular choice.
First, the $SU(9)_U$ GUT will play the role of ETC, because the generated fermion masses will weakly depend on the GUT boson masses 
(here acting as ``ETC" 
boson masses) as
shown in Eq.(\ref{eq5}). This group also contains the standard $SU(5)_{\tiny \textsc{gg}}$ Georgi-Glashow GUT~\cite{gg}. Second, the $SU(4)_{\tt{TC}}$ group contained in the GUT will condense before QCD, generating an appropriate Fermi scale necessary
to break the electroweak group. Note that this choice is based on the most attractive channel (MAC) hypothesis~\cite{cor1,suss}, but it can be relaxed if the GUT breaking can
be promoted at very high energies, where even fundamental scalar bosons may be natural due to the presence of supersymmetry~\cite{su1,su2}. In this case we 
could not neglect the possibility of a small TC group [perhaps $SU(2)$] that condenses at one mass scale larger than the QCD one. Third, the horizontal or family
symmetry is necessary to prevent the first- and second-generation ordinary fermions from coupling to TC. The third fermionic generation will obtain masses
due to diagrams like the one in Fig.1, and will be of order $\lambda_E \mu_{\tt{TC}}$, as described below.

The $SU(9)_U$ group has the following anomaly free fermionic representations~\cite{fra}:
\be 
5\otimes[9,8]_i \oplus 1\otimes [9,2]_i \, ,
\ee
where $[\underline{8}]$ and $[\underline{2}]$ are antisymmetric under $SU(9)_U$, and $i=1,2,3$ is the horizontal index necessary for the replication
of the $SU(3)_H$ families. The decompositions of these representations under $SU(4)_{\tt{TC}}\otimes SU(5)_{\tiny \textsc{gg}}$ are

 \br &&\hspace{-0.5cm}[\bf{9},\bf{2}]_i\nonumber\\
&&(1,10) = \left(\begin{array}{ccccc} 0 & \bar{u_{i}}_{B} & - \bar{u_{i}}_{Y} & -{u_{i}}_{R} & -{d_i}_{R}  \\
-\bar{u_i}_{B} & 0 & \bar{u_i}_{R} & -{u_i}_{Y} &  -{d_{i}}_{Y} \\ \bar{u_{i}}_{Y} & -\bar{u_{i}}_{R} & 0
& -{u_i}_{B} &  -{d_{i}}_{B} \\ {u_i}_{R} & {u_i}_{Y} & {u_i}_{B} & 0 & \bar{e_i}\\
{d_i}_{R} & {d_i}_{Y} & {d_i}_{B} & -\bar{e_{i}}  & 0\end{array}\right)\nonumber\\\nonumber \\ &&(4,5) =
\,\,\,\left(\begin{array}{c} {T_i}_{R} \\ {T_i}_{Y} \\ {T_i}_{B} \\ \bar{L_i}\\ \bar{N_i}
\end{array}\right)_{TC}\,\,\,,\,\,\,(\bar{6},1)= N_{i}\nonumber \\ \nonumber \\
&&\hspace{-0.5cm}[\bf{9},\bf{8}]_i\nonumber\\
&&(1,\bar{5}) =\,\,\, \left(\begin{array}{c} \bar{d_i}_{R} \\ \bar{d_i}_{Y} \\ \bar{d_i}_{B} \\
e_i \\ \nu_{e_i}
\end{array}\right)
\,\,\,\,\,\,(1,\bar{5}) = \left(\begin{array}{c} \bar{X}_{R_{k}} \\ \bar{X}_{Y_{k}} \\ \bar{X}_{B_{k}} \\
E_{k} \\ N_{E_{k}}
\end{array}\right)_i\nonumber\\ \nonumber \\\nonumber\\
&&(\bar{4},1)= \,\,\,\,\,\bar{T_i}_{\varepsilon}, L_i ,{N_i}_{L}. \nonumber \er
\noindent  In the fermionic content of the above model, we identify the usual quarks as $Q = (u ,d)$, while $T$ corresponds to techniquarks and $(L , N)$ to technileptons, where  $\varepsilon = R,Y,B$  is a color index, and $k=1...4$ indicates the generation number of exotic fermions that must be introduced in order to render the model anomaly free.  

The $SU(3)_H$ quantum numbers must be assigned such that the quartet of technifermions that condenses in the MAC of the
product ${\bf \bar{4}\otimes 4}$ belongs to the ${\bf \overline{6}}$ representation of the horizontal group, whereas the QCD quark condensate (generated in the color product 
${\bf \bar{3}\otimes 3}$), is formed in the triplet representation (${\bf 3}$) of $SU(3)_H$. This is nothing else than the horizontal symmetry scheme 
with fundamental scalar bosons proposed in Refs.~\cite{h1,h2,h3,h4}, and it leads to a quark mass matrix in the horizontal group basis of the form
 \br
 m_q =\left(\begin{array}{ccc} 0 & m_1 & 0\\ m_1^* & 0 & 0 \\
0 & 0 & m_3
\end{array}\right),
\label{eq6} \er
\noindent where $m_1$ and $m_3$ indicate the first- and third-generation quark masses.

It is instructive to show the diagrams that lead to the different masses shown in Eq.(\ref{eq6}). For instance, let us assume that
$m_q$ is the mass matrix of charge $2/3$ quarks, where $m_3$ would be related to the top-quark mass. The diagrams responsible for
this mass are shown in Fig.3.
\begin{figure}[h]
\centering
\includegraphics[scale=0.6]{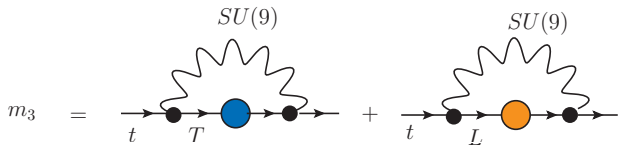} 
\caption[dummy0]{Diagrams contributing to the top-quark mass.}
\label{fig3}
\end{figure}
\par  In this figure the technifermions $T$ and $L$ [that condense in the ${\bf \overline{6}}$ of $SU(3)_H$] give masses to the $t$ quark whose interaction is mediated by one $SU(9)$ gauge boson. Apart from the logarithmic term appearing in Eq.(\ref{eq5}) this mass is
\be
m_3\approx 2 \lambda_9 \mu_{\tt{TC}} \, ,
\label{eq7}
\ee
where we can assume that $\lambda_9 \approx 0.1$, is the product of the $SU(9)$ coupling constant times some Casimir operator eigenvalue, the factor $2$ accounts both diagrams of Fig.3, and $\mu_{\tt{TC}}$ can be assumed to be of $O(1)$TeV. The $SU(9)$ interaction is playing
the role of the ETC interaction. These naive assumptions
will lead to a top-quark mass of approximately $200$GeV. The logarithmic term appearing in Eq.(\ref{eq5}) [and 
neglected in Eq.(\ref{eq7})] slightly decreases the value of our rough estimate.

Note that the first and second charge $2/3$ quarks do not couple directly to the techniquarks
due to the different $SU(3)_H$ quantum numbers, and at this level they remain massless. 

We can now see how the first-generation fermions obtain their masses. In Fig.4 we show the diagrams that are responsible for the $u$-quark mass. 
\begin{figure}[h]
\centering
\includegraphics[scale=0.6]{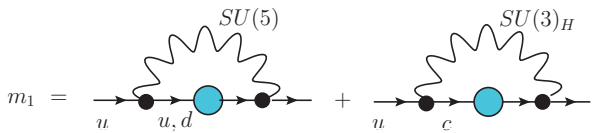} 
\caption[dummy0]{Diagrams contributing to the  light-quarks masses.}
\label{fig4}
\end{figure}
This quark does not couple to techniquarks at leading order, but does couple to other ordinary quark and itself due to the
bosons of the unified theory and the horizontal one. Its mass can be approximated from Eq.(\ref{eq5}) [as we did to obtain Eq.(\ref{eq7})] and
is given by
\be
m_1\approx \lambda_5 \mu_{\tt{QCD}} \, ,
\label{eq8}
\ee
where we can assume naively that the $SU(5)_{\tiny \textsc{gg}}$ factor $\lambda_5 \approx 0.1$ and $\mu_{\tt{QCD}}\approx 200$MeV, which gives 
a mass of order $20$MeV. Here we do not introduce a factor of $2$ in Eq.(\ref{eq8}) due to the presence of the two diagrams in Fig.4, because the $c$-quark condensate (in the second diagram of Fig.4) may be smaller than the $u$ and $d$ condensates\footnote{Note that the self-energy and the condensate values are intimately connected, i.e., one is basically an integral of the other. 
The $c$-quark self-energy appearing in Fig.4 will involve the same type of integral as the $c$-quark condensate. 
It is known that the introduction of heavy quark masses act to diminish
the condensate value or the amount of chiral symmetry breaking ~\cite{x1}. For example, it has been determined for the $s$-quark that 
$\<\bar{s}s\>/\<\bar{u}u\>=0.6\pm 0.1$ ~\cite{x2,x3}. 
In Ref.~\cite{sa5} the same effect of a heavy fermion mass (e.g., fermion loops) was also observed as a factor that lowers the composite Higgs boson mass. Therefore, the second diagram of Fig.4 is
expected to have a smaller effect in the calculation of the first-generation quark masses.}.

In Eqs.(\ref{eq7}) and (\ref{eq8}) we probably overestimated the results when we neglected the logarithmic dependence on
the unified or ``ETC" boson masses. These are very simple calculations. To obtain better estimates we must solve the 
coupled SDE and obtain good fits to the self-energies, which would give us reasonable values for the parameters $\delta_1$ and $\delta_2$ in the
approximate expression of Eq.(\ref{eq4}). It is clear that this is far beyond the scope of this work.

The mass of the second quark generation will necessarily involve the horizontal symmetry, where the coupling to techniquarks will appear
only at two-loop order. The $c$-quark mass will be generated by diagrams like the ones shown in Fig.5, 
\begin{figure}[h]
\centering
\includegraphics[scale=0.55]{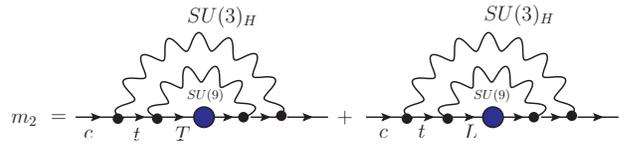} 
\caption[dummy0]{Diagrams contributing to the $c$-quark mass.}
\label{fig5}
\end{figure}
and it is expected to be $1$ order of magnitude below the typical mass of the third quark generation, due to an extra factor $\lambda_{3H}\approx 0.1$ that contains the $SU(3)_H$ coupling constant. In this way, we verify that the horizontal or family symmetry is fundamental to generate a quark mass 
matrix with the Fritzsch texture~\cite{f1,f2}
\br
 m_q =\left(\begin{array}{ccc} 0 & m_1 & 0\\ m_1^* & 0 & m_2 \\
0 & m_2^* & m_3
\end{array}\right),
\label{eq9} \er
 which has several good qualities of the experimentally known quark mass matrix.  
 
Lepton masses will appear in the same way as quark masses. The $\tau$ lepton is the only one that will couple with techniquarks at leading order, due to the appropriate choice of quantum numbers of the horizontal symmetry. As a consequence, the mass matrix  for the leptonic sector is similar to the one described above, although lepton masses should be naturally smaller
than quark masses, because quarks end up coupling to two different condensates and a larger number of diagrams contribute to their masses. It is not difficult to verify
the different number of SDEs between quarks and leptons that can be generated with the Feynman rules of the model described here.

We have not discussed the $SU(9)_U$ and horizontal symmetry breaking, which we just assume to happens at the unification scale $\Lambda_{{}_{SU(9)}}$, 
which can possibly be 
naturally promoted by fundamental scalar bosons. The breaking of the GUT symmetry can also be used to produce a larger splitting in the third fermionic generation.
For instance, if in the $SU(9)_U$ breaking (besides the Standard model interactions and the TC one) we
leave an extra $U(1)$ interaction, we could have quantum numbers such that only the top quark would be allowed to couple to the TC condensate at leading order.
In fact, the splitting ($S_{(t-b)}$) between the $t$ and $b$ quarks
\be
S_{(t-b)}= \frac{m_b}{m_t}\approx \frac{1}{40} \, ,
\ee
is quite large, and it is interesting that the $b$ quark and the $\tau$ lepton could couple at a larger order in the coupling constant [possibly $(\alpha_9^2)$], which could be accomplished by
this remaining $U(1)$ interaction that we referred to above. More sophisticated models in which large fermionic mass splittings and even neutrino masses can be generated
were presented in Refs.~\cite{as1,as2,as3,as4,as5}.

At this point, we hope that we have made clear the necessity of introducing a horizontal or family symmetry. It is necessary to prevent
the first and second generations of ordinary fermions from obtaining large masses that couple to TC at leading order. This symmetry can be a local one, but a
global symmetry is not necessarily discarded. If the family symmetry is local, its breaking can also happen at very high energies and (again) may even be
promoted by fundamental scalars at the GUT or Planck scale, producing feeble effects at lower energies. 

When building a TC model the existence of 
grand unification is also welcome. For example, in the model described here 
a $SU(5)_{\tiny \textsc{gg}}$ gauge boson interaction is fundamental to give the electron a mass, which appears due to the electron coupling to the 
first-generation quark, with exactly the same interaction that may mediate proton decay in the $SU(5)_{\tiny \textsc{gg}}$ theory. There are more diagrams
contributing to the first-generation quark masses than there are for the electron mass, which may explain why leptons are less massive than quarks. 
  
Concerning the possible class of models presented here, it is also clear that a full and precise determination of the mass spectra is quite complex. Once a GUT involving the SM and TC is proposed, we also have to choose
the horizontal symmetry. The coupled SDE of such a model has to be solved by determining all self-energies with their specific infrared and ultraviolet expressions.
Of course, simple estimates can be made by approximating the calculation of each specific fermion mass diagram, by the product of the dynamical mass involved in
the diagram (TC or QCD) with the respective coupling constants and Casimir operator eigenvalues, as performed in Eqs.(\ref{eq7}) and (\ref{eq8}) where a logarithmic
term was neglected.

\section{Scalar mass}

The common lore about theories with a composite scalar boson is that its mass should be of the order of the dynamical mass scale that forms such particle.
This concept is related to the work of Nambu-Jona-Lasinio~\cite{nl} and was also discussed for the $\sigma$ meson in QCD~\cite{ds}, where the scalar composite mass appearing in one strongly interacting theory is given by  
\be
m_\sigma = 2 \mu_{\tt{QCD}}\,\, .
\label{eq10}
\ee
Equation (\ref{eq10}) comes from the fact that at leading order the SDE for the quark propagator is similar to the homogeneous Bethe-Salpeter equation (BSE) 
for a massless pseudoscalar bound state $\Phi_{BS}^P (p,q)|_{q \rightarrow 0}$ (the pion), and a scalar p-wave bound state $\Phi_{BS}^S (p,q)|_{q^2 = 4 \mu^2 }$
[the sigma meson or the $f_0(500)$~\cite{pdg}], i.e.,
\be
\Sigma (p^2) \approx  \Phi_{BS}^P (p,q)|_{q \rightarrow 0} \approx \Phi_{BS}^S (p,q)|_{q^2 = 4 \mu_{\tt{QCD}}^2 }\,\,.
\label{eq11}
\ee
Equation (\ref{eq11}) tells us that in QCD the $\sigma$ meson must have a mass $2\mu_{\tt{QCD}}\approx 500$MeV. In TC we should expect a scalar boson with a mass of $2$TeV, which
is clearly not the case for the observed Higgs boson~\cite{atlas,cms}

There are two subtle points concerning the result of Eq.(\ref{eq10}) and the determination of the scalar composite mass. The first one is that Eq.(\ref{eq10})
was determined using the homogeneous BSE. There is nothing wrong with this. However this gives the right result if the fermionic self-energy that
enters into the BSE is a soft one. When the self-energy decreases slowly [as in Eq.(\ref{eq4})] the scalar mass is modified by the
normalization condition of the inhomogeneous BSE. This modification lowers the composite scalar mass as a consequence
of Eq.(\ref{eq4}). 
The second point about Eq.(\ref{eq10}) that we would like to note is not exactly about the equation itself, but rather about the values of the
dynamical QCD and TC mass scales that arise at such a scale. The QCD dynamical mass scale is usually extracted from the hadronic spectra;
for instance, it is expected to be $1/3$ of the nucleon mass or $1/2$ of the sigma meson mass. However, it is not currently clear how much this spectra
is affected by gluons (or technigluons in the TC case) and mixing among different particles. These points will be discussed in the
following subsections. 

\subsection{Normalization condition and the scalar mass}

The BSE normalization condition in the case of a non-Abelian gauge theory is given by \cite{lane2}
\br
2\imath q_{\mu}= \imath^2\!\!\int d^4\!p\, Tr\left\{{\cal P}(p,p+ q)\left[\frac{\partial}{\partial q^{\mu}}F(p,q)\right]{\cal P}(p, p+ q) \right\}\nonumber \\
-\imath^2\!\!\int d^4\!pd^4\!k \,Tr\left\{{\cal P}(k,k + q)\left[\frac{\partial}{\partial q^{\mu}}K'(p,k,q)\right]{\cal P}(p, p+ q)\right\} \nonumber
\label{eq11a}
\er
where
$$
K'(p,k,q)  = \frac{1}{(2\pi)^4}K(p,k,q)   \,\,\, ,
$$
$$
F(p,q) =  \frac{1}{(2\pi)^4}S^{-1}(p+q) S^{-1}(p) \,\,\, ,
$$ 
where ${\cal P}(p, p + q)$ is a solution of the homogeneous BSE and $K(p,k,q)$ is the fermion-antifermion scattering kernel 
in the ladder approximation. When the internal momentum  $q_{\mu} \rightarrow 0$, the wave function ${\cal P}(p, p + q)$ can be determined only through
the knowledge of the fermionic propagator:
\be 
 {\cal P}(p) = S(p)\gamma_{5}\frac{\Sigma(p)}{F_{\Pi}}S(p) \,\, ,
\ee 
\noindent where $\Sigma (p)$ will describe the technifermion self-energy and it should be noticed  that $F_{\Pi}$ describes the technipion decay constant associated with $n_{d}$  technifermion doublets. If we identify $\Sigma(p^2) \equiv \mu_{\tt{TC}} f(p^2)$ we can write the normalization condition in the rainbow approximation as
\br
&&2i\left(\frac{F_{\pi}}{\mu_{\tt{TC}}}\right)^2 q_{\mu} = \frac{i^2}{(2\pi)^4}\times \nonumber \\
&& \left[\int d^4\!p\, Tr{\Big \{}S(p)f(p)\gamma_{5}S(p )\left[\frac{\partial}{\partial q^{\mu}}S^{-1}(p + q) S^{-1}(p)\right]\right. \nonumber \\ 
&& \left. S(p)f(p)\gamma_{5}S(p){\Big \}} + \frac{i^2}{(2\pi)^4}\int d^4\!pd^4\!k \,Tr{\Big \{} S(k)\right. \nonumber \\
&& \left.f(k)\gamma_{5}S(k)\left[\frac{\partial}{\partial q^{\mu}}K(p,k,q)\right]S(p)f(p)\gamma_{5}S(p){\Big \}}\right]. \nonumber \\
\label{eq11b}
\er 
\par  Equation (\ref{eq11b}) is quite complicated, but it can be separated into two parts: 
\be 
2i\left(\frac{F_{\Pi}}{\mu_{\tt{TC}}}\right)^2 q_{\mu}  = I_\mu^{0} + I_\mu^{K} \,\,\, ,
\label{eq14a}
\ee
corresponding, respectively, to the two integrals on the right-hand side of Eq.(\ref{eq11b}). The fermion propagator given by
$S(p) = {1}/[{\not{\!\!p} - \Sigma(p)}]$ can be written as 
\be 
\frac{\partial}{\partial q^{\mu}}S^{-1}(p + q) =  \gamma_{\mu} -  \frac{\partial}{\partial q^{\mu}}\Sigma(p+q) \,\,\, ,
\ee
\noindent and the term $ \frac{\partial}{\partial q^{\mu}}\Sigma(p+q)$ in the above expression may be written as
\be 
\frac{\partial \Sigma(p+q) }{\partial q^{\mu}} = (p + q)_{\mu} \frac{d\Sigma(Q^2)}{dQ^2}
\ee 
\noindent where $Q^2 = (p + q)_{\mu}(p + q)^\mu$. Considering the angle approximation we transform the term $\frac{d\Sigma(Q^2)}{dQ^2}$ as
\be 
\frac{d\Sigma(Q^2)}{dQ^2} =  \frac{d\Sigma(p^2)}{dp^2}\Theta(p^2 - q^2) + \frac{d\Sigma(q^2)}{dq^2}\Theta(q^2 - p^2)
\ee 
where $\Theta$ is the  Heaviside  step function.  We can finally contract Eq.(\ref{eq14a}) with $q^\mu$ and compute it at $q^2=M_H^2$ in order to obtain
\br 
 M_{H}^2  =  4&&\mu_{\tt{TC}}^2{\Big\{}\frac{n_{f}N_{TC}}{8\pi^2}\int d^2\!p\frac{f^2(p)\Sigma(p)}{(p^2 + \Sigma^2(p))^2} \times \nonumber \\
 && \times \left(-p^2\frac{d\Sigma(p)}{dp^2} \right) \left(\frac{\mu_{\tt{TC}}}{F_{\Pi}}\right)^2  + \nonumber  \\
&& + \,\,I^{K}(q^2 = M_H^2,f(p,k),g_{TC}^2(p,k)) {\Big \}},
\label{eq11c}  
\er 
where $n_f$ is the number of technifermions, $N_{TC}$ is the number of technicolors and $g_{TC}$ is the technicolor coupling constant. 

An expression similar to Eq.(\ref{eq11c}) was already obtained by us in Ref.~\cite{usx}. In that work we just assumed (in a totally \textit{ad hoc} fashion)
a hard momentum behavior for the TC self-energy. The calculation here will differ not only in the origin of the self-energy but also in the approach
we follow to determine the value of $M_H$. Considering the work of Ref.~\cite{us2}
it becomes evident that the behavior of $M_{H}$ is a result that will fundamentally depend on the boundary conditions satisfied by the 
coupled system described in Fig.2. In Eq.(\ref{eq11c}) the UV behavior of the term 
\be 
{\rm \bf (UV)}\,\,\,\,\,\,\,lim_{{}_{{}_{\hspace*{-0.5cm} p^2 \to \Lambda^2}}}\!\!\!-p^2\frac{d\Sigma(p)}{dp^2} ,
\label{newuv} 
\ee
\noindent will be affected by the effective mass generated by the diagrams $(a_2)$, $(a_3)$, and $(a_4)$ in Fig.2. In Ref.~\cite{us2} we verified that
the UV behavior of the term in Eq.(\ref{newuv}) is modified as $\alpha_E$ is different or equal to zero, and we shall comment on this term later.

We compute $M_H$ by numerically solving the
differential coupled equations shown in Eqs.(11) and (12) of Ref.~\cite{us2} , fitting the resulting solutions (all fits with $R^2=0.98$), and inserting the fits into Eq.(\ref{eq11c}). We consider the TC gauge groups $SU(2)_{TC}$, $SU(3)_{TC}$ and $SU(4)_{TC}$, with $n_f=5$ fermions in the fundamental representation, $\mu_{\tt{TC}}=1$TeV, and use the MAC hypothesis to constrain the TC gauge coupling and Casimir eigenvalue. Hereafter, we follow
Refs.~\cite{us1,us2} and use a Casimir eigenvalue $C_E=1$ and gauge coupling constant $\alpha_E = 0.032$, which are quantities related to the ETC gauge theory.

Our results for $M_H$ are shown in Table I, where we can see that the normalization condition lowers the scalar mass by a factor of $O(1/10)$. The results are consistent with those of Ref.~\cite{usx} obtained with the naive assumption of an irregular solution for the TC self-energy. Therefore, the effect
of radiative corrections in coupled SDEs involving a TC theory act in order to produce a scalar composite boson with a mass compatible with that of
the observed Higgs boson.

\begin{table}[htbp]
  \centering
  \begin{tabular}{ccc}\hline \hline
  SU(N) & $n_f$         &         $M_H$(GeV)   \\ \hline
  2 &  5              &          105.3   \\
  3 &  5              &          141.5  \\
  4 &  5             &           148.8     \\
  
     \hline \hline
 \end{tabular}
 \caption{The last column contains the composite scalar mass determined through Eq.(\ref{eq11c}), where we used the TC self-energy obtained by solving the coupled SDE system. The different factors and couplings of the gap equations are described in the text. }
 \label{tbl:IneqFP}
\end{table}

\subsection{Dynamical mass scales and mixing}

The most precise quantity to constrain the dynamical mass scale in the QCD case is the
pion decay constant, which is a function of the quark self-energy. In the TC case the technipion decay constant is 
related to the $W$ and $Z$ gauge boson masses. However, in both cases that quantity depends on the dynamical mass scale as well as the functional
expression for the self-energy. Therefore, we have some freedom in pinpointing the dynamical mass scale. Even the numerical determination of the self-energy
through SDE solutions includes the introduction of a cutoff and specific approximations. We conclude that the calculation of the scalar boson mass
depends on the functional form of the self-energy and on the dynamical mass scale. It is curious that in the past the scalar boson mass
was considered in order to constrain the dynamical mass scale, i.e., in QCD the scalar $\sigma$ meson mass has led to the usual value $\mu_{\tt{QCD}}\approx 250$MeV, which is also approximately
the value of the QCD mass scale ($\Lambda_{\tt{QCD}}$). The problem is that the result of Eq.(\ref{eq10}) is modified not only by the inhomogeneous
BSE condition, butalso by many other effects as we discuss in the following.

The dynamical QCD mass scale is also thought to be related to the nucleon mass, but even this is not certain since we do not know how much gluons
contribute to the nucleon mass~\cite{lorce}. It is also not yet clear how much of the sigma meson mass comes from mixing with heavier
quark-antiquark scalars and with glueballs~\cite{mi1,mi2,mi3,mi4,mi5,mi6}, and the same is true if we just exchange QCD with TC, which means that
the scales $\mu_{\tt{QCD}}$ and $\mu_{\tt{TC}}$ may be smaller than usually thought, leading to a smaller scalar
composite mass (i.e., the $\sigma$ and the ``Higgs" mass). The scalar mass can also be modified by the effect of radiative loop corrections due to
the presence of heavy fermions, as described in Ref.~\cite{sa5}. 

These are not the only effects that modify the scalar mass and lead to a new relation between the scalar mass and the dynamical mass scale. There is still another effect that is intimately related to the type of dynamical
symmetry breaking model that we discussed in the previous section. 

In Sec. II we discussed a model with two composite scalar states responsible for the chiral (and gauge) symmetry breaking: the scalars
belonging to the ${\bf \overline{6}}$ and ${\bf 3}$ representations of the horizontal group formed by technifermions and
quarks, respectively. The different scalars may mix among themselves due to electroweak or other interactions, as already pointed out in Ref.~\cite{us1}. 

An order-of-magnitude estimate of these mixing diagrams is quite lengthy, but the most important
fact is that the scalar coupling to the electroweak bosons is going to be enhanced, when compared to this coupling
calculated when the TC self-energy is soft. Note that this effective coupling happens when scalars and $W$ bosons couple through a ordinary fermion or technifermion
loop. The $W$ coupling to fermions is the SM one, while the scalar composite coupling to ordinary fermions was shown by Carpenter \textit{et al}.~\cite{ca1,ca2} to be
proportional to $\frac{g_w}{2M_W}\Sigma$, where $\Sigma$ is the fermionic self-energy, which now is a slowly decreasing function of momentum and enhances 
the effective coupling. If we denote a composite scalar by $\phi$, it is possible to show that the $\phi\phi WW$ effective coupling will 
be proportional to~\cite{us3}
\be
\Gamma_{\phi\phi WW}\propto \frac{g_W^4\delta^{ab}}{M_W^2}\frac{g^{\mu\nu}}{32\pi^2}\int dq^2 \frac{\Sigma_\phi^2}{q^2},
\label{eq12}
\ee
where $\Sigma_\phi$ has to be substituted by the TC or QCD self-energy depending on which fermion is involved in the composite scalar. Of course, the
complete calculation of the mixing diagrams is quite model dependent, but, as commented in Ref.~\cite{us1}, the origin of this mixing is another
way to see how a full Fritzsch matrix pattern of fermion masses can be generated in the type of model that we are proposing here. It is due to 
this type of coupling that the second-generation fermion masses are generated in models with fundamental scalar bosons~\cite{h1,h2,h3,h4}. Finally, in
the context where all SM symmetry breaking is promoted by composite scalars we cannot even say how much of the $\sigma$ [r $f_0 (500)$] meson mass is due to a possible mixing with a composite Higgs boson.

\section{Pseudo-Goldstone bosons}

In the condensation of the $SU(4)_{\tt{TC}}$ group a large number of Goldstone bosons are formed. Even if we consider other TC groups, only
three of the Goldstone bosons are absorbed in the SM gauge breaking, and regardless of the theory we may end up with several light composite
states resulting from the chiral symmetry breaking of the strong sector. 

These pseudo-Goldstone bosons (or technipions) in the model 
of Sec. II may have different quantum numbers.
They may be colored bosons $~\bar{Q}\gamma_5\lambda^a Q$, where $\lambda^a$ is a color group generator, charged bosons $~\bar{L}\gamma_5 Q$ and
neutral pseudo-Goldstone bosons $~\bar{N}\gamma_5 N$. These bosons receive masses through radiative corrections, and we will verify that,
as a consequence of the logarithmic TC self-energy, they will be heavier than usually thought, which is desired in view of the
stringent limits on light technipions~\cite{scs1}.

In Ref.~\cite{us1} we briefly commented that the technipion masses ($m_\Pi$) are enhanced in comparison with models where the TC self-energy does 
not have the form of Eq.(\ref{eq4}). One of the arguments is quite simple: the technifermions obtain an effective mass ($m_F$) of several GeV 
through diagrams ($a_3$) and ($a_4$) of Fig.2. Note that in our case the condensation effect is not soft, and the calculation of these diagrams will result in
a mass that is not different from those of the third ordinary fermionic family. In particular, in our model there will be several contributions
to these types of diagrams. Even the neutral technifermion $N$ will receive contributions from TC condensation mediated by the electroweak $Z$ boson,
and from QCD condensation due to $SU(9)$ GUT bosons. These masses, apart from small logarithmic terms, will be roughly of order
\be
m_F \approx \sum_{i}\lambda_i \mu_{\tt{TC}} \, ,
\label{eq13}
\ee
where $\lambda_i$ represents the product of some coupling constant times Casimir operator eigenvalue contained in any diagram of the type ($a_3$) or ($a_4$)
contributing to the technifermion mass. For the colored and charged technifermions we cannot even discard a mass as heavy or higher than the top-quark
mass. These masses will generate rather heavy technipions as can be verified using the Gell-Mann-Oakes-Renner relation
\be
m_\Pi^2 \approx m_F \frac{\<{ \bar{\psi}_T}\psi_T\>}{2F_{\Pi}^2} \, ,
\label{eq14}
\ee
where $\<{ \bar{\psi}_T}\psi_T\>$ is the TC condensate and $F_{\Pi}$ is the technipion decay constant. With $m_F$ of order of several GeV and standard values
for the condensate and technipion decay constant the technipion masses turn out to be of order of $100$ GeV or higher, as discussed in Ref.~\cite{us1}.

Another way to see that technipion masses are enhanced through the calculation of a diagram that was already shown in Ref.~\cite{us1}
(see Fig.4 of that reference). Any radiative boson exchange within a technipion modifying its mass will necessarily involve the technipion vertex 
connecting it to technifermions ($\Gamma_{\Pi F}$). However this vertex is proportional to the technipion wave function $\Phi_{BS}^\Pi (p,q)$,
which at leading order is also related to the TC self-energy as
\be
\left.\Phi_{BS}^\Pi (p,q)\right|_{q\rightarrow 0} \approx \Sigma_T (p^2) \, ,
\label{eq15}
\ee
which is responsible for an enhancement of this radiative correction. An order-of-magnitude calculation of such a diagram was presented in Ref.~\cite{us1},
and we will comment later on the phenomenology of technipions with masses that are not very different from that of the Higgs boson.

\section{TC condensate}

In the previous section and throughout this work we have commented about the different condensates (TC and QCD), and it is interesting to make a connection between 
the several studies about the TC condensate value based on walking TC~\cite{yama} and the one we are discussing here. The TC condensate at one high energy 
scale $\Lambda$ is related to its value at another scale $\mu$ by
\be 
\<{ \bar{\psi}_T}\psi_T\>_{\Lambda} = Z^{-1}_m\<{ \bar{\psi}_T}\psi_T\>_\mu \, ,
\ee
where $Z^{-1}_m$ is a renormalization constant which is given by 
$$
Z^{-1}_m \sim \left(\frac{\Lambda}{\mu}\right)^{\gamma_m} \, ,
$$
where $\gamma_m$ is the condensate operator anomalous dimension. 

It is possible to compare the condensate values for a theory where the anomalous  dimension is perturbative and small at high energy, 
i.e. $\gamma_m \to 0$ and the one with a nontrivial large anomalous dimension, for instance, in the extreme walking case where $\gamma_m \to 2$.
We can define the following ratio that measures the difference between condensates in the walking and nonwalking regimes:
\be
R_w = \frac{\<{ \bar{\psi}_T}\psi_T\>_{\Lambda}^{\gamma_m \to 2}}{\<{ \bar{\psi}_T}\psi_T\>_{\Lambda}^{\gamma_m \to 0}} \, ,
\label{rw}
\ee
Considering these extreme cases this ratio is proportional to
\be
\left. R_w\right|_{\gamma_m \to 2} \approx \left( \frac{\Lambda}{\mu}\right)^{2} \, ,
\label{rw2}
\ee
and this expression serves as an indicator of how much the theory is modified by the nontrivial anomalous dimension. This kind of relation can also be used to 
verify how radiative corrections appearing in Fig.2 change the TC behavior.

The UV boundary conditions of the differential TC gap equations modified by the radiative corrections (as can be seen in Ref.~\cite{us2}) are given by
\be 
\left. p^2\frac{d\Sigma(p)}{dp^2}\right|_{\Lambda \to \infty} = -a\int^{\Lambda^2}_0 dk^2\frac{\Sigma(k)}{k^2 + \Sigma^2(p)} \, ,
\ee
where $a$ is a factor involving the gauge coupling constant and Casimir operator eigenvalue related to the interaction that induces the radiative correction [e.g., constants related to one of the diagrams $(a_2)$, $(a_3)$ or
$(a_4)$ in Fig.2]. On the other hand, we recall that in an $SU(N)$ gauge theory the condensate can be represented by
\be 
\<{ \bar{\psi}_T}\psi_T\>_{\Lambda} = -\frac{N}{4\pi^2}\int^{\Lambda^2}_0 dk^2\frac{\Sigma(k)}{k^2 + \Sigma^2(k)} \, .
\ee
These relations allow us to redefine the ratio shown in Eq.(\ref{rw}) where the condensate values are determined with and without radiative corrections, i.e., when they
are calculated with the coupled SDE system ($\alpha_E \neq 0$ ) and with the values of the isolated condensates ($\alpha_E = 0$),
\be
R_w^{rad.cor.} = \frac{\<{\bar{\psi}_T}\psi_T\>^{\alpha_E \neq 0}_{\Lambda}}{\<{\bar{\psi}_T}\psi_T\>^{\alpha_E =0}_{\Lambda  }} 
\approx 
\frac{\left. p^2\frac{d\Sigma(p)}{dp^2}\right|^{\alpha_E\neq 0}_{\Lambda \rightarrow\infty} }{\left. p^2\frac{d\Sigma(p)}{dp^2}\right|^{\alpha_E=0}_{\Lambda \rightarrow \infty}} \, .
\label{rwcor}
\ee 

We computed Eq.(\ref{rwcor}) by considering the solutions of the coupled and isolated SDE system in the case of the $SU(3)$ TC group, with 
$\mu =1$ TeV, $\alpha_E =0.032$,  $\alpha_{{_{TC}}} = 0.87$ and $C_{{}_{TC}}  = 4/3 $. The self-energies were obtained in terms of
the variable $x = p^2/\mu^2$ for each ETC scale $M_E$, and the condensates were integrated from $x=10^2$ up to the UV cutoff $ x_\Lambda= \Lambda^2/\mu^2 \sim 10^7 $. 
The ratio $R_w^{rad.cor.}$ was fitted with $R^2=0.999$ in the form $a_1[ln(M^2_E/\mu^2)]^{a_2}$ and the result is
\be
R_w^{rad.cor.} \propto 7.87 \times 10^6[ln(M^2_E/\mu^2)]^{-4.3} \, .
\label{eq14g}
\ee 
If we consider the value of our cutoff ($\Lambda^2/\mu^2 = 10^7$), we can verify that the effect of the radiative correction is not exactly that of
the extreme walking case shown in Eq.(\ref{rw2}), but it is still quite large. We again see that the effect of radiative corrections is not that
different from the effect of the \textit{ad hoc} four-fermion interactions determined by Takeuchi~\cite{takeuchi}. Moreover, if we compute
the generated quark mass ($m_Q$) as a function of the TC condensate we obtain
\be 
m_Q \approx \frac{\<{\bar{\psi}_T}\psi_T\>^{\alpha_E \neq 0}_{\Lambda}}{\Lambda^2} \approx C [ln(M^2_E/\mu^2)]^{-\kappa_2}, 
\label{eq14j}
\ee 
where the constant $C \sim O(\mu)$. This behavior is consistent with that of Eq.(\ref{eq5}).

\section{Experimental constraints}

\subsection{$S$ parameter}

The $S$ parameter provides an important test for new physics beyond the Standard Model~\cite{pt}. This parameter can be described by the absorptive
part of the vector-vector minus axial-vector-axial-vector vacuum polarization in the following form in the case of a TC model with
new composite vector and axial-vector mesons with masses $M_V$ and $M_A$ and respective decay constants $F_V$ and $F_A$~\cite{pt}:
\be
S=4\int_0^\infty \frac{ds}{s} Im \overline{\Pi}(s)=4\pi \left[ \frac{F_V^2}{M_V^2}-\frac{F_A^2}{M_A^2} \right] \, .
\label{eqss}
\ee

An interesting analysis of the $S$ parameter in TC theories was performed in Ref.~\cite{asan} with the use of the Weinberg sum rules,
where the case of a conformal theory was considered. In our case, we have a TC model which is just a scaled QCD theory,
with effective masses due to the different SDE contributions shown in Fig.2, besides its dynamical mass of $O(1)$ TeV. 
There is no reason to expect modifications of Eq.(\ref{eqss}) for this type of theory, as well as the simple extension to TC of the first and  second Weinberg sum rules,
which are respectively
\be 
F_V^2 - F_A^2 = F_\Pi^2 \, ,
\label{1sm}
\ee
and
\be
F_V^2 M_V^2 - F_A^2 M_A^2 =0 \, ,
\label{2sm}
\ee
which lead to
\be
S=4\pi F_\Pi^2 \left[ \frac{1}{M_V^2}+\frac{1}{M_A^2} \right] \, .
\label{sfin}
\ee

We can also apply the result of vector meson dominance to Eq.(\ref{sfin})~\cite{wei2}, implying that $M_A^2 = 2 M_V^2$. This relation
is not exact even in QCD, but by considering it we are at most overestimating the $S$ parameter, which is now be given by
\be
S\approx \frac{6\pi F_\Pi^2}{M_V^2} \, .
\label{s2}
\ee
The TC technipion decay constant is usually assumed to be $F_\Pi \approx 246$GeV. 

To determine the value of $S$ shown in Eq.(\ref{s2}) we must have one
estimate of the vector-meson mass. It should be remembered that the vector-boson mass is quite large only due to the spin-spin part
of the hyperfine interactions. We can determine the vector-boson mass by using the hyperfine splitting calculation performed in the
heavy quarkonium context in Ref.~\cite{ei}
\be
M(^3S_1)-M(^1S_0)\approx \frac{8}{9} {\bar{g}}^2(0) \frac{|\psi (0)|^2}{\mu^2} \, ,
\label{eqvc}
\ee
where $M(^3S_1)$ and $M(^1S_0)$ describe the masses of vector and scalar lighter bosons, respectively. In Eq.(\ref{eqvc}), $|\psi (0)|^2$ is the meson wave 
function at the origin, describing a vector boson formed by techniquarks with dynamical mass $\mu_{\tt{TC}}$. Equation (\ref{eqvc}) seems to be reasonable
even when the vector-boson constituents are light~\cite{sc}.

We make the following assumptions: 1) The TC theory has an infrared frozen coupling constant ${\bar{g}}^2(0)/4\pi \approx 0.5$, whose value can be similar to several determinations of this quantity in the QCD case (see, for instance, Ref.~\cite{usf}), 2) The lightest TC scalar boson has the same mass as the Higgs
boson found at the LHC, i.e., $M(^1S_0)=125$GeV, 3) The wave function is approximated by $|\psi (0)|^2 \approx \mu_{\tt{TC}}^3\approx 1$TeV$^3$, consistent
with the other BSE wave functions proportional to the dynamical fermion mass (see Eq.(\ref{eq11})). As a consequence, we obtain
a vector-boson mass $M_V\approx 5.71$TeV, leading to
\be
S \approx  0.035 \, ,
\label{sfu}
\ee
whose value has probably been overestimated but is still consistent with the experimental data ($S=0.02\pm 0.07$)~\cite{pdg}.

\subsection{Horizontal symmetry}

A necessary condition for the type of model that we are proposing here is the presence of the horizontal (or family) symmetry. This symmetry can be local, and
it is only necessary to enforce the connection between the TC sector and the third ordinary fermionic generation, i.e., the $t$ and $b$ quarks, the $\tau$, and
its neutrino. This symmetry in general leads to flavor violations at an undesirable level; however, in the scheme proposed here the masses of the horizontal
gauge bosons can be quite heavy, affecting only logarithmic corrections to the fermion masses, and not producing significant tree-level reactions
that may be severely constrained by the experimental data. On the other hand there are hints of $B$ decay anomalies~\cite{b1,b2,b3,b4,b5} which, if confirmed, could also set a mass scale for our horizontal symmetry.

One of the anomalies in $B$ decays appears in the measurement of the ratio between the branching fractions of the processes $B^0 \rightarrow K^{*0}\mu^+\mu^-$ and
$B^0 \rightarrow K^{*0}e^+e^-$, which in the small dilepton invariant mass region is given by
\be
R (K^*)=\frac{B^0 \rightarrow K^{*0}\mu^+\mu^-}{B^0 \rightarrow K^{*0}e^+e^-}= 0.66 {\substack{+0.11 \\ -0.07}}\pm 0.03 \, ,
\label{bdec}
\ee
which is around $2.2$ standard deviations away from the SM expectation.

If such deviation is confirmed in the future, it could be explained by a current-current interaction described by the following effective Lagrangian:
\be
L_h \propto \alpha_h \frac{\lambda_{bs}C^{\mu\mu}}{M_h^2} (\overline{s}\gamma_\nu P_L b)(\overline{\mu}\gamma^\nu \mu) \, ,
\label{lb}
\ee
where $\alpha_h$ is the horizontal gauge coupling, $\lambda_{bs}$ are mixing angles, $M_h$ is the horizontal gauge boson mass, and $C^{\mu\mu}$ is a Wilson
coefficient. If we naively assume the results of the $SU(3)_h$ horizontal model of Ref.~\cite{alo} for these several constants, we can roughly estimate that
$M_h$ should be greater than $10$TeV. However, this is only a guess because (as said repeatedly in the previous sections) the horizontal gauge boson
can be quite heavy, and this scale can be set to these masses only if the anomalies remain discrepant with the SM expectation. Otherwise, the
dependence on the factor $1/M_h^2$ in all observables of this kind will lessen experimental constraints originated from horizontal symmetries.

There are other possible flavor-changing neutral currents induced by the horizontal symmetry. For instance, the effective Lagrangian
\be
L_h \propto \alpha_h \frac{\lambda_{sd}}{M_h^2} (\overline{s}_L\gamma_\nu d_L)(\overline{s}_R\gamma^\nu d_R) \, ,
\label{sd}
\ee
is induced by one-gauge-boson exchange and contributes to the $K^0 - \bar{K}^0$ transition, which for $\lambda  \approx 1/20$ requires $M_h \geq 200$TeV ~\cite{die}.
This contribution can be easily evaded in our type of model simply by increasing the horizontal gauge boson mass scale, which will not affect the mechanism
of ordinary fermion mass generation. Therefore, a careful scrutiny of the gauge symmetry
breaking of the horizontal group will only be necessary if the $B$ decay anomaly is confirmed.

\subsection{Technipion masses} 

The LHC collaborations already have enough data to constrain the existence of light technipions~\cite{scs1}. Due to the fact that the technifermions acquire masses
of $O(100)$ GeV, the resulting pseudo-Goldstone bosons [i.e., those generated in the chiral breaking of the $SU(4)_{\tt{TC}}$ TC gauge group discussed in Sec. II]
may be heavier than the SM Higgs boson. Moreover, due to the choice of the horizontal symmetry quantum numbers the technipions will mainly couple to the third ordinary fermionic family,
i.e., $t$ and $b$ quarks and the $\tau$ lepton, in such a way that may easily evade the limits found in Ref.~\cite{scs1} obtained from data on the SM Higgs boson decaying
into $\gamma \gamma$ and $\tau^+ \tau^-$. 

The colored and charged technipions will be quite heavy and are produced along with $t$ and $b$ quarks. In the case of the decay into $b$ quarks the branching ratio
may be reduced by a possible small coupling between this quark and the technipion, which will happen through the exchange of a rather heavy gauge boson, and 
their signal could easily be buried in the background. This leaves us with the lightest technipions, which should be the 
neutral ones ($\bar{N}\gamma_5 N$). In this case a neutral technipion may be produced through vector-boson fusion and decay through the weak $ZZ$ channel. 

The discussion of the TC condensate in Sec. V can be used to estimate the neutral technipion mass ($m_\Pi$) in a different way than in Ref.~\cite{us1}. 
As considered in Eq.(\ref{eq14j}) the neutral technifermion mass ($m_N$) in terms of the TC condensate generated by diagram $(a_4)$ of Fig.2 is given by
\be
m_N \sim \frac{\<{\bar{\psi}_T}\psi_T\>^{\alpha_E \neq 0}_{\Lambda}}{\Lambda^2} \, .
\ee
The above equation together with Eq.(\ref{eq14}) leads to the following estimate of the neutral technipion mass
\be
 m_\Pi^2 \approx  \frac{(\<{\bar{\psi}_T}\psi_T\>^{\alpha_E \neq 0}_{\Lambda})^2}{2F_{\Pi}^2\Lambda^2} \, .
\ee
Assuming $SU(3)_{{TC}}$ as the TC gauge group, $\<{\bar{\psi}_T}\psi_T\>^{\alpha_E =0} \sim \mu^3$ with $\mu = 1$ TeV, 
$R_w^{rad.cor.} \approx  7.87 \times 10^6[ln(M^2_E/\mu^2)]^{-4.3}$ defined and appearing in Eqs.(\ref{rwcor}) and (\ref{eq14g}), we obtain
\be 
 m_\Pi \sim  160 \,\,\,  GeV  ,
\ee
which is a rough estimate for the smallest pseudo-Goldstone mass of our type of model, which has not yet been eliminated by the LHC data~\cite{scs1}.

The fact that in our type of model the technifermions couple preferentially to the third fermionic family and obtain a large effective mass due ETC interactions, and that their other couplings to ordinary fermions are always diminished by the exchange of a very heavy horizontal or GUT gauge boson makes the search for pseudo-Goldstone signals quite difficult. The main hope for detecting technipions may be the resonant production of the lightest neutral technipion and its decay into neutral weak bosons.

\section{Scalar boson trilinear coupling}

As already pointed out many years ago~\cite{ebo}, the measurement of the Higgs boson trilinear coupling is fundamental to determining the nature of
this particle. If the Higgs boson is a composite particle its trilinear coupling may deviate from the SM value of a fundamental scalar boson,
and its measurement can even provide a signal of the underlying theory forming the composite state~\cite{doff}.
 
In TC or any composite scalar model the scalar trilinear coupling is determined through its coupling to fermions.
Using Ward identities, we can show that the couplings of the scalar boson to fermions are \cite{ca2}
\be 
G^{\sl a} (p+q,p) = -\imath \frac{g_{W}}{2M_{W}}
\left[\tau^{\sl a}\Sigma(p)P_R - \Sigma(p+q)\tau^{\sl a} P_L \right]
\label{fsc}
\ee
where $P_{R,L} = \frac{1}{2} (1 \pm \gamma_5 )$, $\tau^{\sl a}$ is a $SU(2)$
matrix, and $\Sigma$ is the fermionic self-energy in weak-isodoublet
space. As in Ref.\cite{ca2}, we assume that the scalar composite Higgs
boson coupling to the fermionic self-energy is saturated by the
top quark. We also do not differentiate between the two fermion momenta $p$ and $p+q$ since, in all situations of interest, 
$\Sigma(p+q)\approx \Sigma(p)$. Therefore, the coupling between a composite Higgs boson and fermions at large momenta is given by  
\be 
\lambda_{{}_{Hff}}(p)\equiv G(p,p) \sim -\frac{g_{W}}{2M_{W}}\Sigma(p^2).
\label{fh}
\ee  

The trilinear coupling of the composite scalar boson is determined by the diagram shown in Fig.6.  
\begin{figure}[ht]
\begin{center}
\includegraphics[scale=0.5]{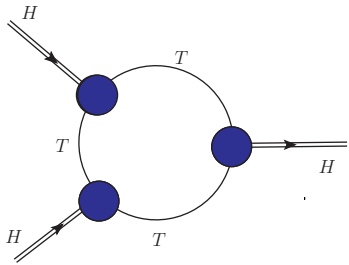}
\vspace{0.3cm}
\caption{The dominant contribution to the trilinear scalar coupling. The blobs in this figure represent the coupling of the composite scalar boson to fermions. The double lines represent 
the composite scalar boson.}
\label{fig6}
\end{center}
\end{figure}
Assuming that the coupling of the scalar boson to the fermions is given by Eq.(\ref{fh}), we find that
\be 
\lambda_{{}_{3H}} = \frac{3g^3_{W}}{64\pi^2}\left(\frac{3n_{F}}{M^3_{W}}\right)\int^{M^2_E}_{0}\frac{\Sigma^4(p^2)p^4dp^2}{(p^2 + \Sigma^2(p^2))^3}.
\label{tri} 
\ee 
\noindent where  $n_{F}$ is the number of technifermions included in the model. 

The SM trilinear scalar coupling value, according to the normalization of Ref.~\cite{malt}, is
\be 
\lambda_{SM} = \frac{M^2_H}{2v^2}.
\label{norml}
\ee
Combined with the above normalization, the trilinear coupling of Eq.(\ref{tri}) leads to the following scalar trilinear coupling $\lambda$: 
\be 
\lambda =  \frac{1}{6v}\lambda_{{}_{3H}} .
\label{nor2}
\ee

Considering Eqs.(\ref{tri}) and (\ref{nor2}), $v=F_{{}_{\Pi}}$, and the relation 
$$
M^2_{W} = \frac{g^2_W F^2_{{}_{\Pi}}}{4}
$$
we obtain for the trilinear coupling
\be 
\lambda = \frac{1}{16\pi^2}\left(\frac{3n_{F}}{F^4_{{}_{\Pi}}}\right)\int^{M^2_E}_{0}\frac{\Sigma^4(p^2)p^4dp^2}{(p^2 + \Sigma^2(p^2))^3},
\label{tri3} 
\ee
which is the trilinear scalar composite coupling that can be compared to the SM coupling of Eq.(\ref{norml}).

\begin{figure}[t]
\begin{center}
\includegraphics[scale=0.45]{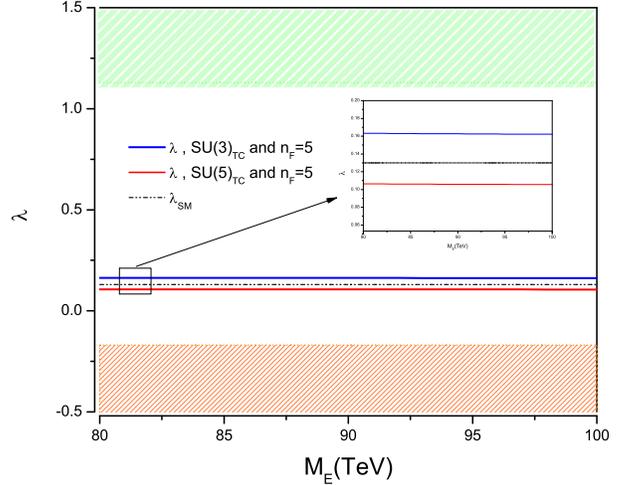}
\vspace{-2cm}
\caption{Experimental limits on the scalar boson trilinear coupling, and curves of the trilinear coupling value (\ref{tri3})) in the case of a composite scalar boson. }
\label{fig7}
\end{center}
\end{figure}

Using the results for the TC self-energy obtained in Ref.~\cite{us2} and Sec. III, which is dominated by diagrams ($a_1$) and ($a_4$) of Fig.2,  we compute the trilinear coupling presented in Eq.(\ref{tri3}). A comparison of the trilinear composite coupling with the SM one is shown in Fig.7. The composite trilinear coupling does differ from the SM one, 
but only a small amount. In Fig. 7 we also show the current LHC limits on this coupling obtained in Ref.~\cite{malt} from the $(b\bar{b}\gamma\gamma)$ signal, whose values are 
 $\lambda < -1.3\lambda_{SM} = -0.169$ (red region) and  $\lambda  > 8.7\lambda_{SM} = 1.13$ (green region). Figure 7 remind us that the actual result
for the scalar trilinear coupling does vary with $M_E$, and this variation should appear when the coupled gap equations are solved taking into account the
running of the ETC gauge coupling constant. Of course, this will introduce only a small variation in the curves of that figure. 
The white region is not excluded yet, and this large region shows how difficult it is to differentiate one composite scalar boson from a fundamental one by just observing the specific coupling.

\section{Conclusions}

In Refs.~\cite{us1,us2} we called attention to the fact that the self-energies of strongly interacting theories are modified when we consider coupled SDEs including radiative
corrections. The effect of the radiative corrections is not very different from the \textit{ad hoc} introduction of effective four-fermion interactions, as verified many
years ago by Takeuchi~\cite{takeuchi}, and it leads to self-energies that decrease logarithmically with the momentum. This effect was reviewed in the Introduction of this work, where it was made clear that the usual TC model building has to be modified, where the ordinary fermion mass hierarchy is not related to different ETC gauge boson masses.

The presence of a horizontal symmetry is mandatory in the type of models envisaged in Sec. II. This symmetry is necessary to give masses to only the third generation of
ordinary fermions at leading order. The model discussed in Sec. II is based on the non-Abelian gauge group structure $SU(9)_U \otimes SU(3)_H$, where the $SU(9)_U$
group contains the SM, an $SU(5)_{\tiny \textsc{gg}}$ Georgy-Glashow GUT~\cite{gg}, and a $SU(4)_{\tt{TC}}$ group. The $SU(3)_H$ horizontal symmetry was introduced in
such a way that their fermionic quantum numbers allow only the third fermionic generation to be coupled to the technifermions. The other fermions remain massless at
leading order. However, the first-generation fermions obtain their masses due to the coupling with QCD, which also has a slowly decreasing self-energy. This is the most
interesting fact of our model: the different fermionic mass scales are dictated by the different strong interactions present in the model! We have shown some of the diagrams that
generate the different masses, and made rough estimates of their masses. We believe that a large number of theories can be built along the lines of the model of
Sec. II. Precise determinations of fermion masses in this type of model will demand a lengthy determination of SDE coupled equations, where different self-energies
can be fitted by equations like Eq.(\ref{eq5}).  

The fact that the ETC interactions can be pushed to very high energies apparently seems to open
a path for a plethora of TC models capable of describing the ordinary fermionic mass spectra. 
The determination of fermion masses will involve a delicate balance of different gauge group
theories for TC, ETC (or GUT), and horizontal symmetry. The ordinary fermion mass matrix calculation will involve the knowledge
of specific Casimir eigenvalues, which will depend on the different fermionic representations of the different gauge groups.
It will also involve the different coupling constant values of these theories at different scales, and the far more demanding
solutions of the coupled system of Schwinger-Dyson equations even with a minimum of approximations. Therefore, while
a new frontier arise, generic combination of gauge theories and 
respective fermionic representations will not be able to explain the known fermionic spectra, meaning that an enormous engineering effort
will be necessary for a \textit{precise} calculation of ordinary fermion masses.

In Sec, III we discussed how the composite scalar boson may have a mass lighter than the characteristic mass scale of the theory that forms the composite particle.
This could explain how the observed Higgs boson mass, if composite, is smaller than the Fermi mass scale. Perhaps the most important factor regarding the mass value of the scalar 
composite resides in the normalization
condition of the inhomogeneous BSE, which has to be taken into account when the self-energy is hard and not decaying as $1/p^2$. The normalization condition,
as shown by the results presented in Table I, is enough to lower the scalar mass by a factor of $1/10$. However, we have listed many other effects
that may also lower the scalar composite mass.

Section IV contains a brief discussion about pseudo-Goldstone boson masses. It is just a complementary discussion to the one already presented in Refs.~\cite{us1,us2}, 
indicating that their masses should be of the order of or higher than that of the observed Higgs boson. Moreover, the pseudo-Goldstone bosons couple at leading order
only to the third-generation fermions, which is another fact that will complicate their experimental observation.

In Sec. V we computed the TC condensate in the coupled SDE scenario. This calculation serves as a comparison with the enhancement that appears in the TC
condensate in walking TC theories. Although the mechanism is totally different, i.e., here the gauge theory is just a running theory, there is also one enhancement in the
condensates as a result of a logarithmically decreasing self-energy with the momentum. Again, it is possible to verify that the effect is not qualitatively different
from the \textit{ad hoc} inclusion of a four-fermion interaction, which is replaced by genuine radiative corrections of known interactions.

In Sec. VI we commented on possible experimental constraints on this type of model. The main point is that the ETC gauge boson masses may be pushed to very high
energies and unnatural flavor-changing events will be absent. The $S$ parameter will be of the expected order, and should not differ from the case of TC as a scaled QCD theory. Complementing the discussion of Sec. IV with what was presented in Sec. V, we estimated pseudo-Goldstone masses and verified that they cannot yet be seen
at the LHC according the analysis of Ref.~\cite{scs1}.

In Sec. VII we computed the trilinear scalar coupling and verified that a signal of compositeness is far from being observed with the present data~\cite{malt},
and this coupling does not differ by a large amount from the SM value in the case of a fundamental scalar boson. Finally, we may say that in the scenario
presented in this work there is a possibility that the SM gauge symmetry breaking promoted dynamically by composite scalar bosons is still alive. 

\section*{Acknowledgments}

This research  was  partially supported by the Conselho Nacional de Desenvolvimento Cient\'{\i}fico e Tecnol\'ogico (CNPq)
under the grants 302663/2016-9 (A.D.) and 302884/2014 (A.A.N.).

\appendix

\end{document}